\documentclass[pre,twocolumn,showpacs]{revtex4}

\usepackage{amssymb}
\usepackage{amsfonts}
\usepackage{amstext}
\usepackage{graphicx}
\usepackage{amscd}
\usepackage{amsmath}

\newcommand{\tr}{\mathrm{tr}}

\begin{document}

\title{Decoherence, relaxation and chaos in a kicked-spin ensemble -- Schr\"odinger's cat kicked by Arnold's cat}

\author{David Viennot}
\author{Lucile Aubourg}
\affiliation{Institut UTINAM (CNRS UMR 6213, Universit\'e de Franche-Comt\'e, Observatoire de Besan\c con), 41bis Avenue de l'Observatoire, BP1615, 25010 Besan\c con cedex, France.}

\begin{abstract}
We study the dynamics of a quantum spin ensemble controlled by trains of ultrashort pulses. To model disturbances of the kicks, we consider that the spins are submitted to different kick trains which follow regular, random, stochastic or chaotic dynamical processes. These are described by dynamical systems on a torus. We study the quantum decoherence and the population relaxation of the spin ensemble induced by these classical dynamical processes disturbing the kick trains. For chaotic kick trains we show that the decoherence and the relaxation processes exhibit a signature of chaos directly related to the Lyapunov exponents of the dynamical system. This signature is a \textit{horizon of coherence}, i.e. a preliminary duration without decoherence followed by a rapid decoherence process.
\end{abstract}


\maketitle

\section{Introduction}
Real quantum systems are never isolated, interactions with its environment induce quantum decoherence \cite{breuer}, i.e. transitions from quantum state superpositions into incoherent classical mixtures of eigenstates. In order to study this process, spin baths \cite{lages,gedik,lages2,rossini,zhou,castanino,xu,brox} or quantum kicked tops \cite{wang,jacquod,znidaric,ghose} are interesting systems. Spin baths can be studied by NMR experiments and can be viewed as assemblies of qubits (a qubit is an unit of quantum information in quantum computing). These applications need to study the control of spin baths \cite{lapert,zhang,lapert2}, but the decoherence processes can drastically decrease the efficiency of the control. Previous studies concerning decoherence of spin baths focused on decoherence induced by spin-spin interactions inner the bath (the spin bath being itself considered as an environment for one of its spins). In this paper we study decoherence processes induced by disturbances of the control caused by a classical environment. Since it is a simple but efficient control method \cite{sugny,dion,sugny2}, we focus on a control by a train of ultrashort pulses (kicks). In order to enlighten the role of classical control disturbances in decoherence processes and to avoid other decoherence causes, we consider a spin bath without any spin-spin interaction. Strictly speaking we consider then a simple spin ensemble but which could be considered as a spin bath for one spin chosen in the ensemble. Since we do not consider any spin-spin interaction, the central system (the chosen spin) does not directly interact with its bath. Nevertheless the control disturbance induces a statistical state distribution on the spin ensemble. The chosen spin inherit a mixed state (a density matrix) from the statistical distribution of the ensemble. In this sense, we can consider the spin ensemble as a model of a simple spin bath. We study different disturbance processes that we model by classical dynamical systems which can be regular, random, stochastic or chaotic.\\
\indent The role of the chaos in quantum decoherence processes is a subject of debate \cite{lages,gedik,lages2,rossini,zhou,castanino,xu,brox,wang,jacquod,znidaric,ghose}. But in these studies, the considered chaos is associated with the inner dynamics of the spin bath. In this paper we consider chaos associated with disturbance processes on the control of the spin bath. We show that a signature of this chaos can be observed in the decoherence processes and that a Lyapunov exponent analysis is relevant to describe this phenomenon.\\
\indent This paper is organized as follows. Section 2 is devoted to a presentation of the considered model, of its dynamics and of the classical dynamical systems modeling disturbances. Section 3 analyses the decoherence processes occuring in the kicked spin ensemble for non chaotic dynamical systems. Section 4 talkes about the chaotic case, its Lyapunov exponent and its entropy analysis. Finally section 5 summarizes the properties of the different processes. Throughout this paper, we consider with a particular interest the relation between the classical dynamical system (disturbed kick train) with the quantum dynamical system (spin ensemble) controlled by the classical one. The possibility that some dynamical properties are transmit from classical systems to quantum systems is not well known. In particular we focus on the disorder transmission between the two systems and on the role of the sensitivity to initial conditions (SIC) in chaotic cases.

\section{Dynamics of kicked spin ensembles}
\begin{figure}
\includegraphics[width=8.5cm]{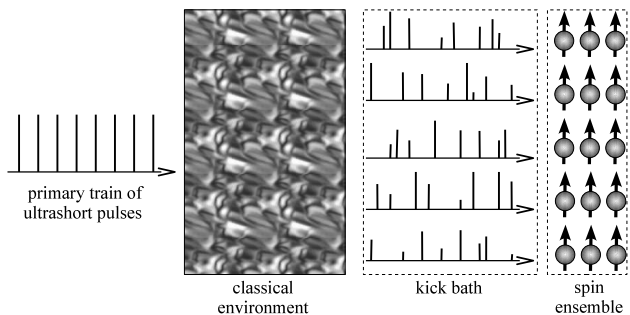}
\caption{\label{kickedspinbath} Schematic representation of a quantum spin ensemble controlled by a disturbed train of ultrashort pulses. The set of kick trains issued from the distrubance constitutes a kind of ``classical kick bath''.}
\end{figure}

We consider an ensemble of $N$ spins without spin-spin interaction. A constant and uniform magnetic field $\vec B$ is applied on the spin ensemble inducing an energy level splitting by Zeeman effect. We denote by $\frac{\hbar \omega_1}{2}$ the energy splitting. At the initial time $t=0$ the bath is completely coherent, i.e. all the spins are in the same quantum state $|\psi_0\rangle = \alpha |\uparrow \rangle+ \beta |\downarrow \rangle$ ($|\alpha|^2+|\beta|^2=1$ with $\alpha,\beta\not=0$ -- $|\psi_0\rangle$ is a ``Schr\"odinger's cat state'' -- ). For $t>0$ the bath is submitted to a train of ultrashort pulses kicking the spins. We suppose that a classical environment disturbes the pulses such that each spin ``views'' a different train (fig. \ref{kickedspinbath}).
We denote by $\omega_0 = \frac{2\pi}{T}$ the kick frequency of the primary train. We suppose that the classical environment can attenuate kick strengths and can delay kicks. We denote by $\lambda_n^{(i)}$ and by $\tau_n^{(i)}$ the strength and the delay of the $n$-th kick on the $i$-th spin of the bath. Let ${H_0 = \frac{\hbar \omega_1}{2} |\downarrow \rangle \langle \downarrow|}$ be the quantum Hamiltonian of a single spin with only the Zeeman effect (where we have removed a constant value without significance). The quantum hamiltonian of the $i$-th kicked spin is
\begin{equation}
H^{(i)}(t) = H_0 + \hbar W \sum_{n \in \mathbb N} \lambda_n^{(i)} \delta\left(t-nT+\tau_n^{(i)} \right)
\end{equation}
where $\delta(t)$ is the Dirac distribution and where the kick operator $W$ is a rank one projector: $W = |w \rangle \langle w|$ with the kick direction ${|w \rangle = \cos \vartheta |\uparrow \rangle + \sin \vartheta |\downarrow \rangle}$ (for the sake of simplicity we do not consider a relative phase between the two components of $|w \rangle$). By considering the reduced time $\theta = \frac{2\pi t}{T} = \omega_0 t$ we have
\begin{equation}
\label{dynamics}
H^{(i)}(\theta) = H_0 + \hbar\omega_0 W \sum_{n \in \mathbb N} \lambda_n^{(i)} \delta\left(\theta-2n\pi+\varphi_n^{(i)} \right)
\end{equation}
with the angular delay $\varphi_n^{(i)} = \omega_0 \tau_n^{(i)}$. The $n$-th monodromy operator (the evolution operator from $t=\frac{2n\pi}{\omega_0}$ to $\frac{2(n+1)\pi}{\omega_0}$) is \cite{viennot}
\begin{equation}
\label{monodromy}
U_n^{(i)} = e^{-\frac{\imath H_0}{\hbar \omega_0} (2\pi - \varphi_n^{(i)})} (id+(e^{-\imath \lambda_n^{(i)}}-1)W)e^{-\frac{\imath H_0}{\hbar \omega_0} \varphi_n^{(i)}}
\end{equation}
We see that the monodromy operator is $2\pi$-periodic with respect to the kick strength; $\lambda_n^{(i)}$ is then defined modulo $2\pi$ from the viewpoint of the quantum system. Thus the strength-delay pair $(\lambda,\varphi)$ defines a point on a torus $\mathbb T^2$ which plays the role of a classical phase space for the kick train.\\
Let $|\psi_n^{(i)} \rangle$ be the state of the $i$-th spin at time $t=nT$ ($|\psi_n^{(i)} \rangle$ represents the ``stroboscopic'' evolution of the spin). By definition of the monodromy operator we have
\begin{equation}
|\psi_{n+1}^{(i)} \rangle = U_n^{(i)} |\psi_n^{(i)} \rangle
\end{equation}
The density matrix of the spin ensemble is then
\begin{equation}
\rho_n = \frac{1}{N} \sum_{i=1}^N |\psi_n^{(i)}\rangle \langle \psi_n^{(i)}|
\end{equation}
$\rho_n$ represents the mixed state of the ``average'' spin of the bath \cite{breuer,bengtsson}. It encodes two fundamental informations:
\begin{itemize}
\item the populations $\langle \uparrow| \rho_n |\uparrow \rangle$ and $\langle \downarrow |\rho_n|\downarrow \rangle$ which are the average occupation probabilities of the states $|\uparrow \rangle$ and $|\downarrow \rangle$;
\item the coherence $|\langle \uparrow| \rho_n |\downarrow \rangle|$ which measures the coherence of the bath (the coherence is maximal if all the spins are in the same quantum state -- which can be a quantum superposition --, and is zero if the bath is a classical mixture of spins with some in the state $|\uparrow \rangle$ and others in the state $|\downarrow\rangle$).
\end{itemize}

We consider the kick trains as classical dynamical systems on the torus $\mathbb T^2$. Let $\Phi$ be the (discrete time) classical flow of these dynamical systems.
\begin{equation}
\left( \begin{array}{c} \lambda_n^{(i)} \\ \varphi_n^{(i)} \end{array} \right) = \Phi^n \left(\begin{array}{c} \lambda_0^{(i)} \\ \varphi_0^{(i)} \end{array} \right)
\end{equation}
In the following, we will consider the different kick baths defined by the following flows:
\begin{enumerate}
\item \textit{Stationary bath} defined by the stationary flow
\begin{equation}
\Phi \left( \begin{array}{c} \lambda \\ \varphi \end{array} \right) =  \left( \begin{array}{c} \lambda \\ \varphi \end{array} \right)
\end{equation}
\item \textit{Drifting bath} defined by the flow
\begin{equation}
\Phi \left( \begin{array}{c} \lambda \\ \varphi \end{array} \right) =  \left( \begin{array}{c} \lambda+ \frac{2\pi}{a} \mod 2 \pi \\ \varphi + \frac{2\pi b}{a} \mod 2 \pi \end{array} \right)
\end{equation}
where $a,b \in \mathbb R \setminus \mathbb Q$. The orbit of $(\lambda_0,\varphi_0)$ by $\Phi$ is dense on $\mathbb T^2$.
\item \textit{Microcanonical bath} defined by a flow $\Phi$ consisting to random variables on $\mathbb T^2$ with the uniform probability measure:
\begin{equation}
d\mu(\lambda,\varphi) = \frac{d\lambda d\varphi}{4\pi^2}
\end{equation}
where $\mu$ is the Haar probability measure on $\mathbb T^2$.
\item \textit{Markovian bath} defined by a stochastic flow $\Phi$ consisting to random variables on $\mathbb T^2$ with the following probability measure:
\begin{equation}
d\nu_n(\lambda,\varphi) = \frac{d\lambda d\varphi}{\sqrt{2\pi \sigma}} e^{- \frac{1}{2\sigma} \left((\lambda-\lambda_{n-1})^2+(\varphi-\varphi_{n-1})^2\right)}
\end{equation}
This process is a discrete time Wiener process (a random walk) corresponding to a Brownian motion on $\mathbb T^2$ with average step equal to $\sigma>0$.
\item \textit{Chaotic bath} defined by a conservative chaotic flow $\Phi$ as for example the Arnold's cat map:
\begin{equation}
\Phi \left( \begin{array}{c} \lambda \\ \varphi \end{array} \right) = \left(\begin{array}{cc} 1 & 1 \\ 1 & 2 \end{array} \right)  \left( \begin{array}{c} \lambda \\ \varphi \end{array} \right) \begin{array}{c} \mod 2 \pi \\ \mod 2 \pi \end{array}
\end{equation}
This flow is chaotic and mixing (and then ergodic).
\end{enumerate}
In addition to the flow, kick baths are defined also by the initial distribution of the first kicks $\{ (\lambda_0^{(i)},\varphi_0^{(i)}) \}_{i=1,...,N}$. These first kicks are randomly chosen in $[\lambda_*,\lambda_*+d_0] \times [\varphi_*,\varphi_*+d_0]$ (with uniform probabilities). $(\lambda_*,\varphi_*)$ can be viewed as the parameters of the primary kick train. The length of the support of the initial distribution (the initial dispersion) $d_0$ is the magnitude of the disturbance on the first kick.\\

Next section studies the dynamics of the spin ensemble from the viewpoint of the decoherence processes for the kick baths 1 to 4. The case of chaotic baths is treated in section 4 (with the Arnold cat map and other hyperbolic automorphisms of the torus).

\section{Decoherence and related processes}
Strictly speaking the decoherence consists to a decrease of the spin ensemble coherence ${|\langle \uparrow|\rho_n|\downarrow \rangle|}$ with $n$. The decoherence process is complete if ${\lim_{n \to + \infty} |\langle \uparrow|\rho_n|\downarrow \rangle|} = 0$, it is then a transition from a coherent superposition of quantum states to an incoherent classical mixture of the two eigenstates. The decoherence is often associated with a relaxation process ${\lim_{n \to + \infty} \langle \uparrow|\rho_n|\uparrow \rangle} = p_{\infty}^{\uparrow}$. When $p_{\infty}^{\uparrow} = \frac{1}{2}$ (independently of $|\psi_0\rangle$) the relaxation consists to a transition to a spin ensemble in the quantum microcanonical distribution (equilibrium of a pseudo-isolated quantum bath). If the decoherence occurs without relaxation process then it is called \textit{pure dephasing} (since it is only induced by relative phases in the quantum states of the spins). We study these two processes for a bath constituted by $N=1000$ spins (the results do not evolve significantly for $N$ larger than this value).

\subsection{Decoherence process}
\begin{figure}
\includegraphics[width=8.5cm]{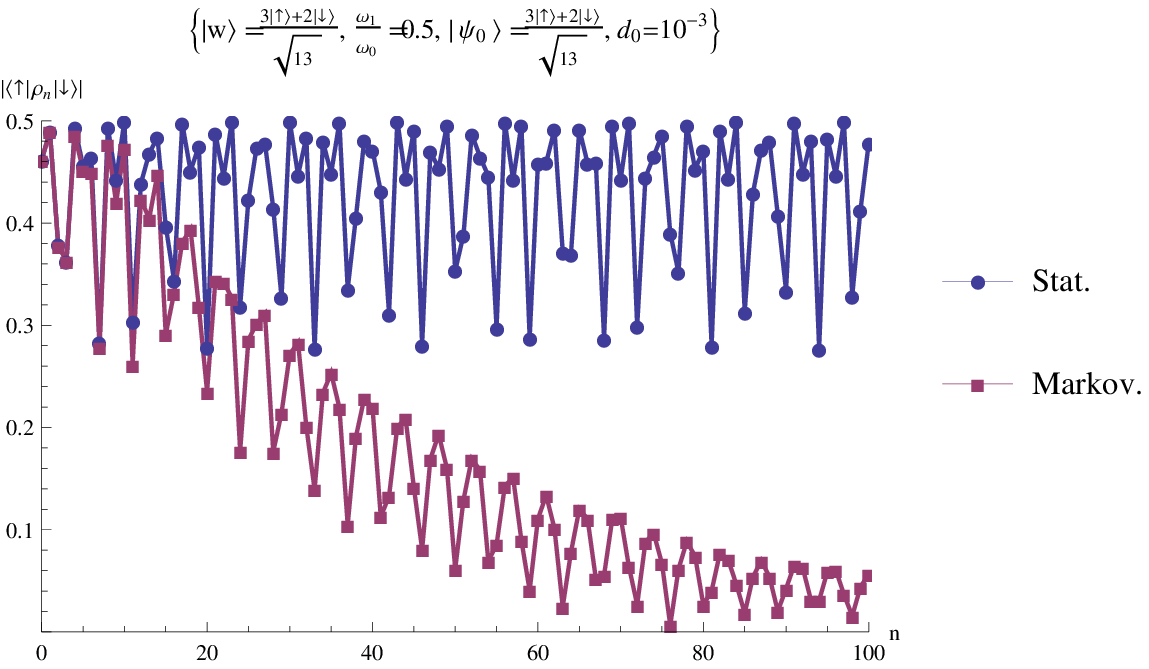}
\includegraphics[width=8.5cm]{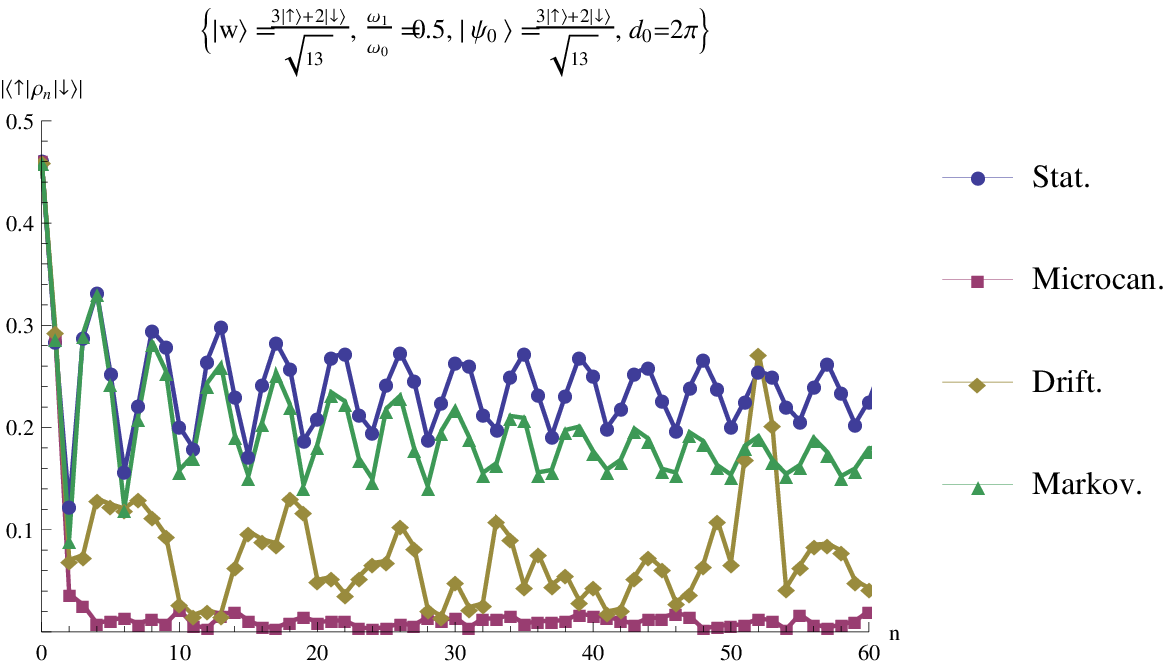}
\caption{\label{coherences} Evolutions of the coherence ${|\langle \uparrow|\rho_n|\downarrow \rangle|}$ of the spin ensemble submitted to a stationary, a drifting, a microcanonical, and a Markovian kick baths (up: for a small dispertion of the first kicks, down: for a large dispertion).}
\end{figure}

\begin{figure}
\includegraphics[width=8.5cm]{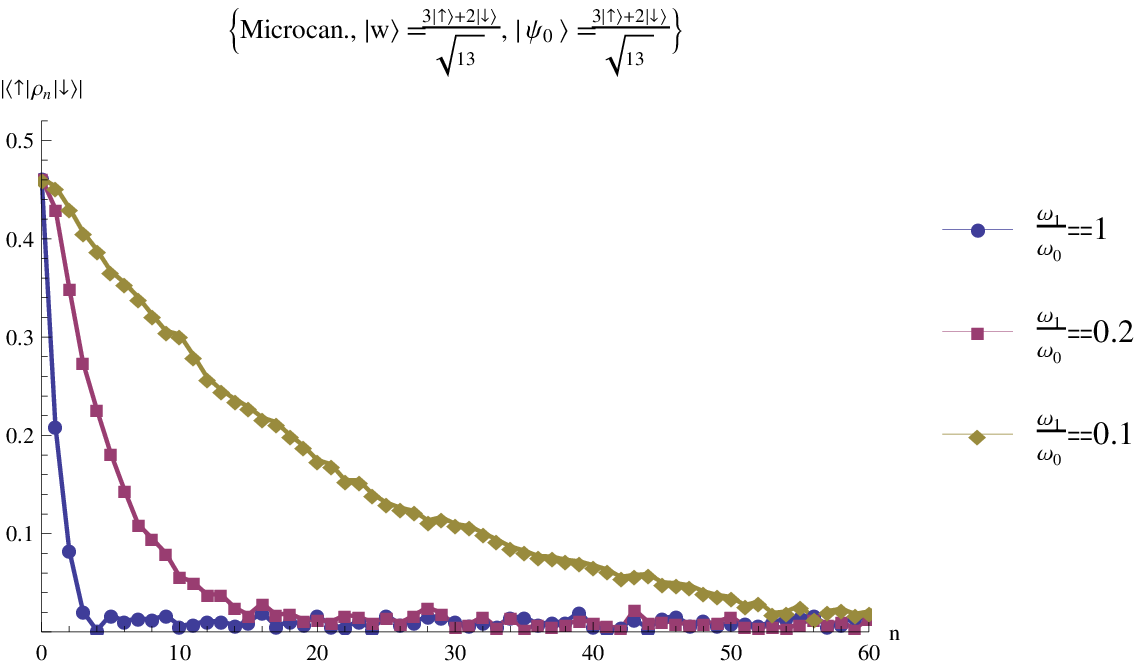}
\includegraphics[width=8.5cm]{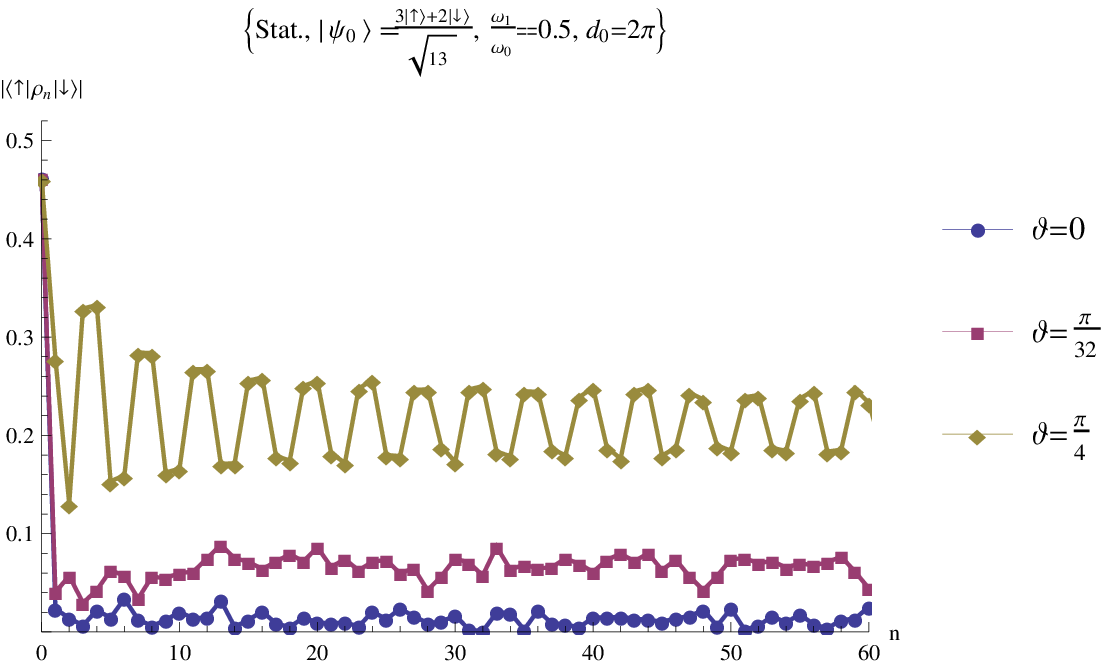}
\caption{\label{para_coherence} Up: Evolutions of the coherence ${|\langle \uparrow|\rho_n|\downarrow \rangle|}$ for a spin ensemble submitted to a microcanonical kick bath  with respect to $\frac{\omega_1}{\omega_0}$. The fall speed of the coherence increases for high values of $\frac{\omega_1}{\omega_0}$.\\
Down: Evolutions of the coherence for a spin ensemble submitted to a stationary kick bath ($d_0 \gg 1$) with respect to $\vartheta$ ($|w \rangle = \cos \vartheta |\uparrow \rangle+\sin \vartheta |\downarrow \rangle$) . $c_{min}$ decreases for $\vartheta$ close to $0$ or $\frac{\pi}{2}$.}
\end{figure}

Fig. \ref{coherences} presents the evolutions of the coherence ${|\langle \uparrow|\rho_n|\downarrow \rangle|}$ for different kick baths.
The decoherence occurs if the kicks are dispersed on $\mathbb T^2$ (microcanonical bath, stationary and drifting bath with a large initial dispersion $d_0$, and Markovian bath with a large initial dispersion or a large average step $\sigma$). The decoherence needs a large dispersion of kicks for the following reason. Suppose that the dispersion of the kicks rests small during the evolution. Each spin is kicked with a similar way than the others. All the spins react then with a similar way, and their states rest approximately equal during the dynamics. The spin ensemble remains then coherent.\\
We see fig. \ref{coherences} that the decoherence is more efficient for an irregular bath in time. Indeed for the microcanonical kick bath we have $\lim_{n \to + \infty} |\langle \uparrow|\rho_n|\downarrow \rangle| = 0$ with a quasi monotonic decrease, whereas for the drifting and the stationary baths (with $d_0 \gg 1$) the decrease comes with large fluctuations. Moreover, for the stationary bath (the more regular example), the decoherence is not complete: $\lim_{n \to + \infty} |\langle \uparrow|\rho_n|\downarrow \rangle| = c_{min} \not= 0$. With $d_0 \gg 1$ the kick bath presents initially a strong disorder, but with the regular evolution the disorder remains constant. Since the disorder of the kick bath does not increase (and is not necessary maximal at time $t=0$), the loss of coherence of the spin ensemble, which consists to a disorder transmission from the kick bath to the spin ensemble, is not optimal.\\
Fig. \ref{para_coherence} shows that the efficiency of the decoherence is strong for a kick direction $|w\rangle$ close to an eigenvector $|\uparrow \rangle$ or $|\downarrow \rangle$ ($\vartheta$ in the neighbourhood of $0$ or $\frac{\pi}{2}$) and for $\frac{\omega_1}{\omega_0} \gg 1$.
For $\omega_1 \ll \omega_0$ the decoherence is inefficient because the proper quantum time of reaction of a spin $\frac{2\pi}{\omega_1}$ is very large in comparison with the time between two kicks $\frac{2\pi}{\omega_0}$. The spins have not the time to evolve between two kicks. The system is kicked so much that it cannot evolve significantly at short-term. The different spin states change then slowly and the loss of coherence is slow. For a kick direction close to an eigenvector, the kicks tend to suppress quantum superpositions (to ``align'' the spins along an eigenvector which is a ``classical direction''). The loss of coherence, which is a loss of pure quantum behaviors, is then naturally favoured in this configuration.

\subsection{Relaxation}
\begin{figure}
\includegraphics[width=8.5cm]{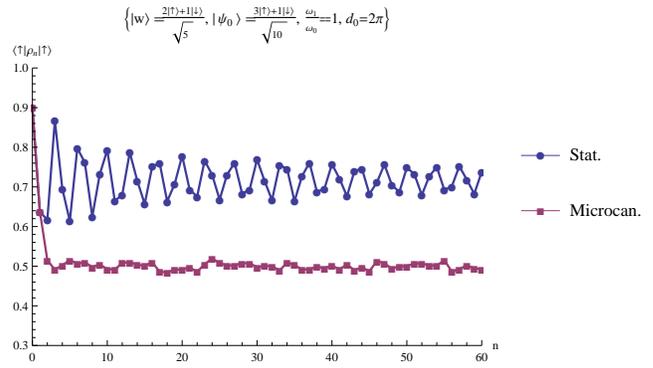}
\caption{\label{populations} Evolutions of the population ${\langle \uparrow|\rho_n|\uparrow \rangle}$ of a spin ensemble submitted to a stationary and a microcanonical kick baths.}
\end{figure}

\begin{figure}
\includegraphics[width=8.5cm]{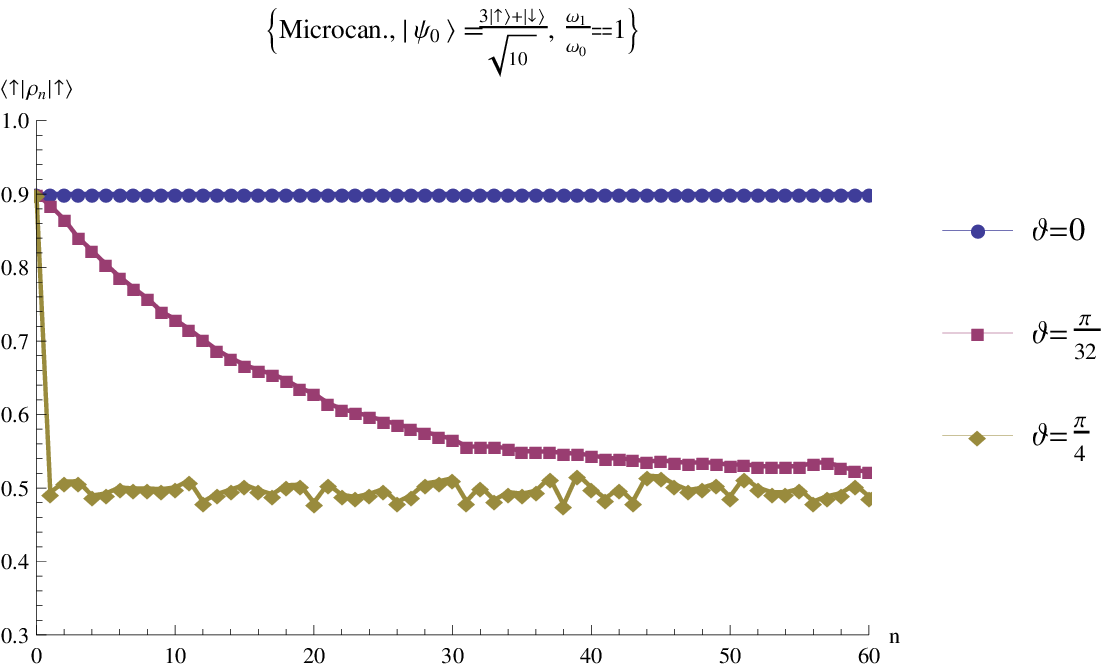}
\includegraphics[width=8.5cm]{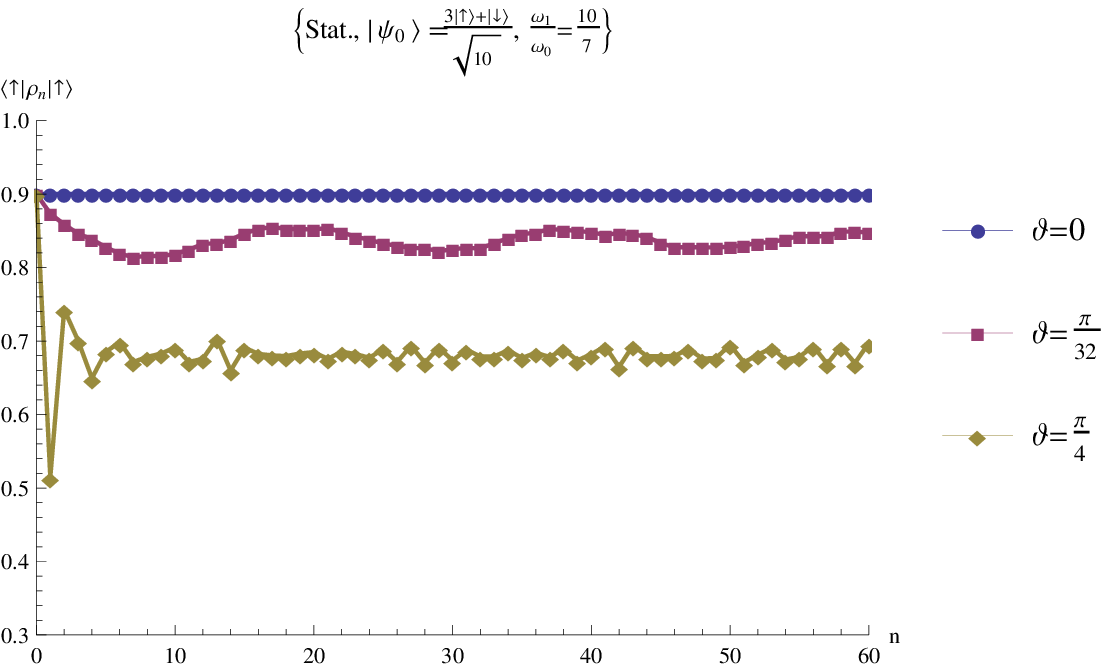}
\caption{\label{para_populations} Up: Evolutions of the population ${\langle \uparrow|\rho_n|\uparrow \rangle}$ for a spin ensemble submitted to a microcanonical kick bath, with respect to $\vartheta$ ($|w \rangle = \cos \vartheta |\uparrow \rangle+\sin \vartheta |\downarrow \rangle$). The speed of the relaxation increases for $\vartheta$ far from $0$ and $\frac{\pi}{2}$.\\
Down: Evolutions of the population for a spin ensemble submitted to a stationary bath ($d_0 \gg 1$) with respect to $\vartheta$. $\lim_{n \to + \infty} {\langle \uparrow|\rho_n|\uparrow \rangle}= p^\uparrow_\infty$ approaches to $\frac{1}{2}$ for $\vartheta$ far from $0$ and $\frac{\pi}{2}$.}
\end{figure}

Fig. \ref{populations} shows the evolutions of the population ${\langle \uparrow|\rho_n|\uparrow \rangle}$ for different kick baths.
The relaxation occurs for dispersed and irregular kicks in the time. The relaxation consists to a loss of the memory of the initial state $|\psi_0\rangle$. With irregular kicks, the disorder in the kick bath is strong and it induces strong increases of the disorder in the spin ensemble. The disorder in the spin ensemble (which can be measured by its entropy) is usually assimilated to a lack of information concerning the spin ensemble \cite{breuer,bengtsson}. The increase of the disorder is then assimilated to a loss of information. The information concerning the memory of the initial state, is lost because the disorder increases with irregular kicks. For the more efficient kick baths we have $\lim_{n \to +\infty} |\langle \uparrow|\rho_n|\uparrow \rangle| = \frac{1}{2}$ (in the same time, the coherence tends to $0$). The spin ensemble is then in the quantum microcanonical distribution which corresponds to the maximal lack of information (the memory of the initial state is completely lost).\\
As for the decoherence process, the relaxation is more efficient for $\frac{\omega_1}{\omega_0} \gg 1$ (if $\omega_1 \leq \omega_0$ the relaxation does not occur for the extreme case of the stationary kick bath). The reason is the same that for the decoherence, for $\omega_1 \leq \omega_0$ the spin system has not the time to react between two kicks. In contrast with the decoherence process, fig. \ref{para_populations} shows that the efficiency of the relaxation grows for kick direction $|w \rangle$ far from the eigenvectors $|\uparrow \rangle$ and $|\downarrow \rangle$.
Indeed, if $|w\rangle = |\uparrow \rangle$ or $|\downarrow \rangle$, then $(|\uparrow \rangle, |\downarrow \rangle)$ are also eigenvectors of $W$ (the kick operator). They are then eigenvectors of the monodromy operator eq. (\ref{monodromy}). The dynamics induces only relative phases between the components $\uparrow$ and $\downarrow$ of the spin wave functions. The populations, which are not sensitive to relative phases, are not modified. For a kick direction along an eigenvector, the decoherence process is a pure dephasing.

\subsection{Population oscillations}
\begin{figure}
\includegraphics[width=6.5cm]{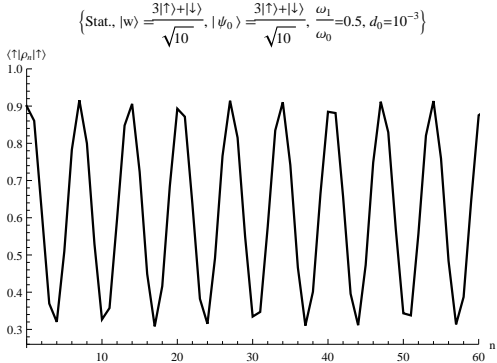}
\caption{\label{fluctuations} Example of fluctuating population $\langle \uparrow|\rho_n|\uparrow \rangle$.}
\end{figure}

\begin{figure}
\includegraphics[width=6.5cm]{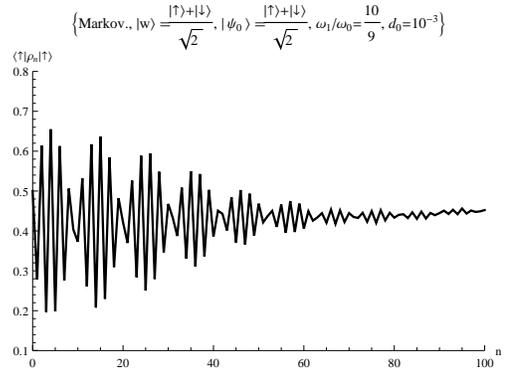}
\caption{\label{oscillations} Oscillations of the population $\langle \uparrow |\rho_n|\uparrow \rangle$ for a Markovian kick bath.}
\end{figure}

For weakly dispersed kicks and/or time regular kicks ($d_0 \ll 1$ in a stationary, a drifting or a Markovian kick bath), the population presents rapid fluctuations (see fig. \ref{fluctuations}).
These fluctuations are generated by Rabi oscillations of the spin states. The kicks must be weakly dispersed because for dispersed kicks the Rabi oscillations are very different from a spin to another one, and they disappear with the average on the bath. The fluctuation frequency is then $\omega_1$ (the Rabi frequency). We can note that the amplitude of these fluctuations decreases for $|w \rangle$ close to an eigenvector, because for $|w \rangle = |\uparrow \rangle$ or $|\downarrow \rangle$, $(|\uparrow \rangle,|\downarrow \rangle)$ become eigenvectors of the evolution operator eq. (\ref{monodromy}).\\
For the Markovian case, these fluctuations become neat oscillations which can be damped and can present beats, fig. \ref{oscillations}. 
The damping increases if the ``average step of the random walk'' $\sigma$ increases. The damping is induced by the increase with time of the dispersion of the kicks (which is quick for large $\sigma$). The kicks becoming dispersed, the coherence of the bath decreases and the Rabi oscillations are killed by the average on the bath.

\subsection{Population jumps and coherence falls}
\begin{figure}
\includegraphics[width=8.5cm]{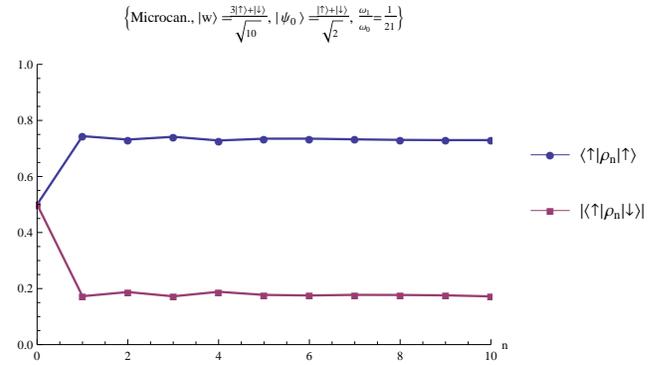}
\caption{\label{jumps} An example of a population jump with a coherence fall (the decoherence and the relaxation processes are slow with respect to the duration represented in this figure).}
\end{figure}

\begin{figure}
\includegraphics[width=8.5cm]{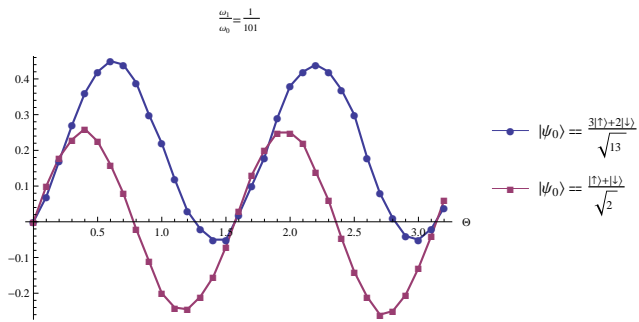}
\caption{\label{jump_ampl} Amplitude of the population jump with respect to $\Theta = \arccos \langle w|\psi_0 \rangle$. The curves can be interpolated by sinusoidal curves.}
\end{figure}

In some cases, we can observe short-term evolutions similar to the decoherence and to the relaxation processes (which are long-term processes), fig. \ref{jumps}.
The population ``jumps'' to another value different from the initial one, and in the same time the coherence falls. This phenomenon is clearly apparent for initially strongly dispersed but time regular kicks (for the drifting bath with $d_0 \gg 1$) and for $\frac{\omega_1}{\omega_0} \ll 1$. Numerical tests show that the direction of the population jump is in the ``direction'' defined by the kick direction $|w \rangle$. The kick tends to ``align'' the spin in its direction. Fig. \ref{jump_ampl} shows the amplitude of the jump with respect to $|\psi_0 \rangle$ and $|w \rangle$.
This phenomenon needs that $\omega_1 \ll \omega_0$ so that the spins do not evolve significantly between two kicks and do not then lose the alignment. The condition concerning strong dispersion of the initial kicks ensures only that the jump is not hide by fluctuations issued from Rabi oscillations. This dispersion is responsible of the coherence fall. The spins being kicked with different strengths, the state changes in the direction of $|w\rangle$ are with different amplitudes for the different spins. This induces a loss of coherence in the spin ensemble.

\section{Chaotic kick baths}
\subsection{Decoherence with the Arnold's cat map}
\begin{figure}
\includegraphics[width=6.5cm]{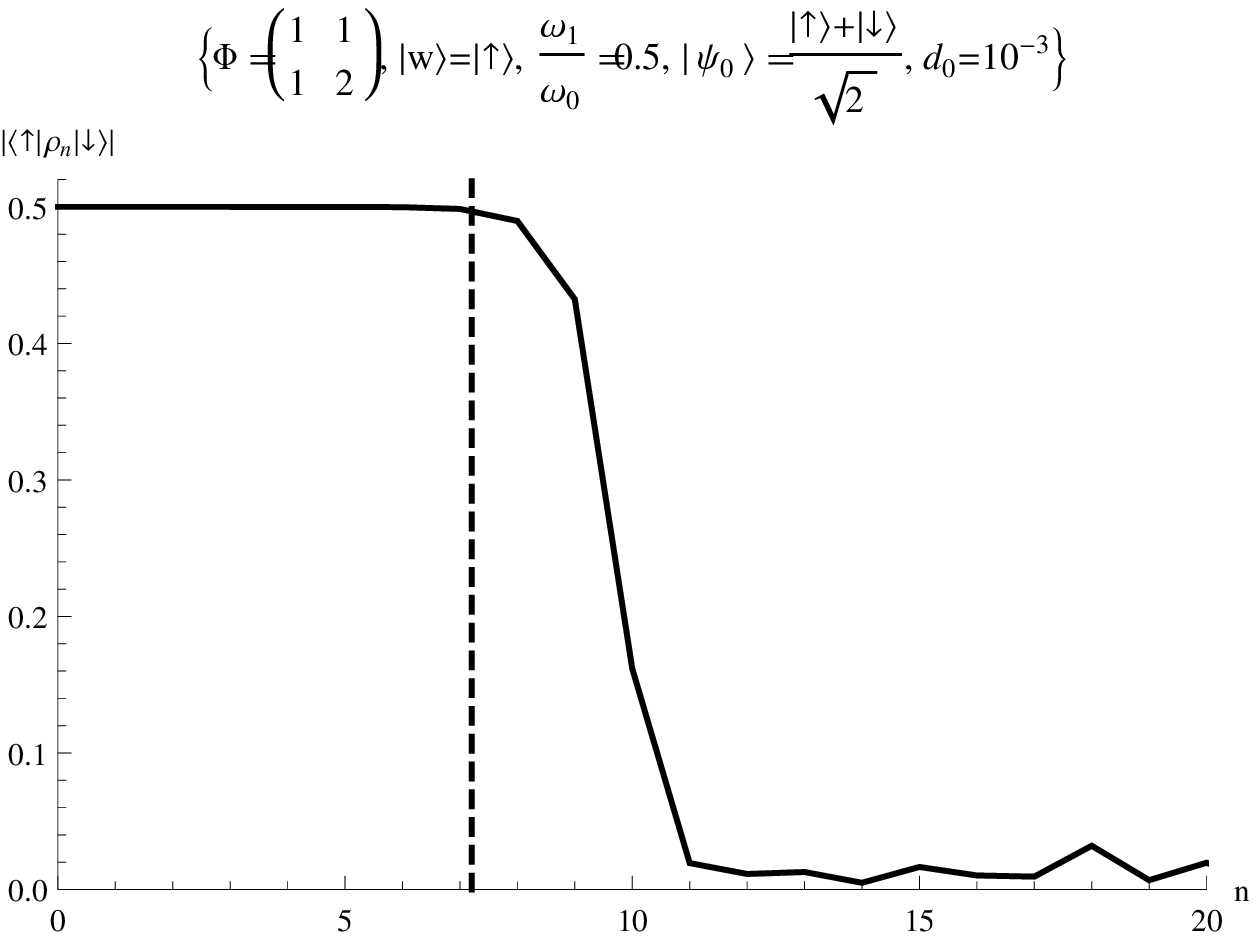}\\
\includegraphics[width=6.5cm]{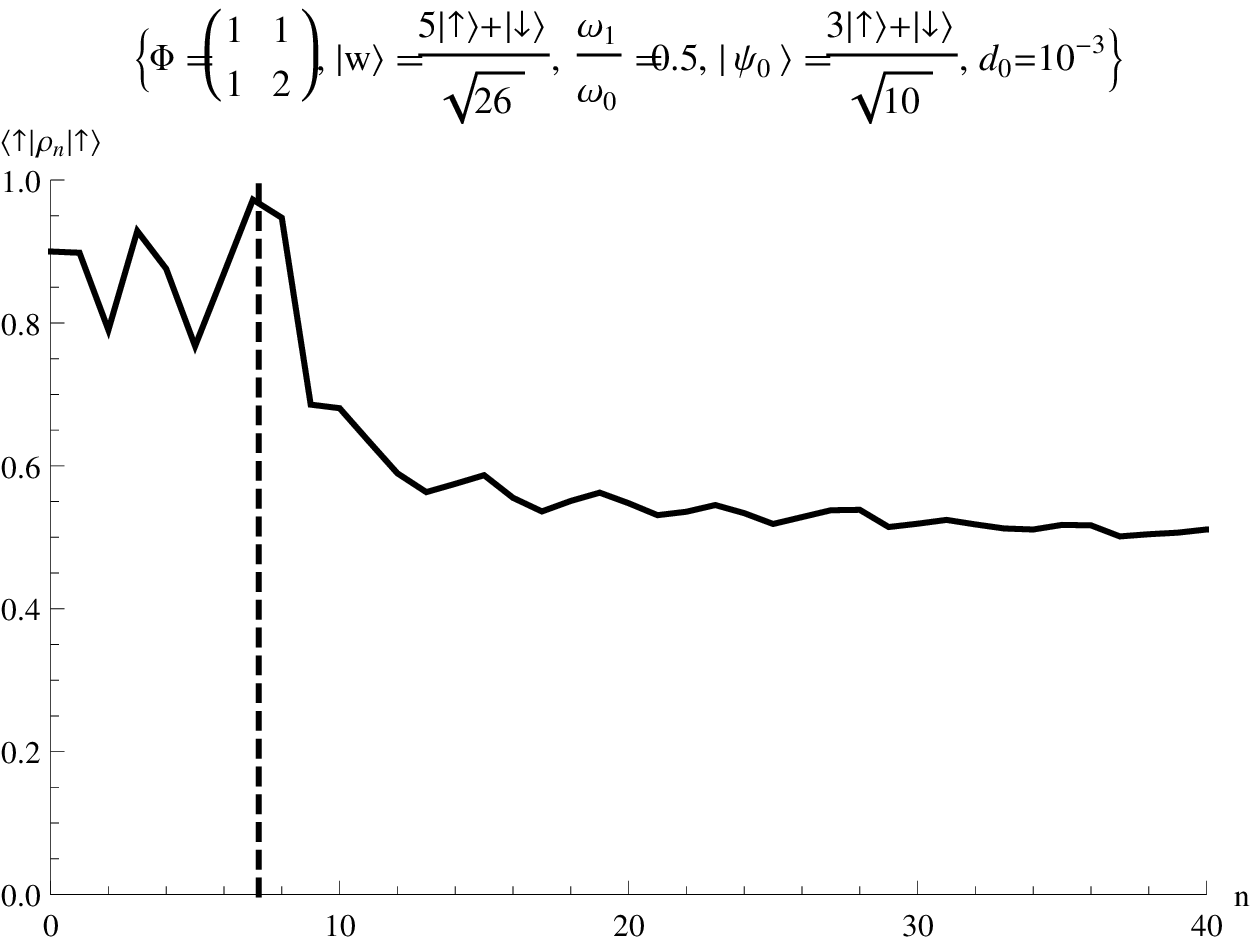}
\caption{\label{arnold_cat} Evolutions of the coherence ${|\langle \uparrow|\rho_n|\downarrow \rangle|}$ and of the population $\langle \uparrow|\rho_n|\uparrow \rangle$ of a spin ensemble submitted to a chaotic kick bath of which the dynamics is defined by the Arnold's cat map. The vertical dashed lines represent the horizon of coherence calculated with eq. (\ref{nstar_th}).}
\end{figure}

We consider now a chaotic kick bath defined by the Arnold's cat map $\Phi = \left(\begin{array}{cc} 1 & 1 \\ 1 & 2 \end{array} \right)$ on $\mathbb T^2$ with a small initial dispersion $d_0=10^{-3}$. Fig. \ref{arnold_cat} shows the evolution of the coherence of spin ensemble $|\langle \uparrow|\rho_n|\downarrow \rangle|$ and the evolution of the population $\langle \uparrow|\rho_n|\uparrow \rangle$.
During a period of approximately 7 iterations, the spin ensemble remains coherent. In the same time the population fluctuates arround its initial value. The behavior of the spin ensemble is then similar to the behavior of a spin ensemble submitted to a regular kick bath (stationary or drifting bath) with a small initial dispersion. But after the period of 7 iterations, the coherence falls rapidly to zero and the population relaxes rapidly to $\frac{1}{2}$. This behavior is similar to the behavior of a spin ensemble submitted to an irregular kick bath with a large initial dispersion (microcanonical bath). These results are confirmed by dynamics with other parameters. The spin ensemble submitted to a chaotic kick bath is the only one which presents two distinct dynamical behaviors quickly succeeding each other. The system presents then a \textit{horizon of coherence} (equal to 7 kicks in fig. \ref{arnold_cat}). Until this horizon of coherence, the spin ensemble is not subject to the decoherence processes, after this horizon the decoherence and the relaxation processes dominate the evolution. This behavior is directly related to the chaotic property of the Arnold's cat map and more precisely to the sensitivity to initial conditions (SIC). At the begining of the dynamics, the kicks of the different spins are approximately identical (the dispersion of the kicks is small, the kick bath is strongly ordered). No disorder can be transmitted to the spin ensemble which remains coherent. But the SIC separates quickly two orbits initially closed, and then increases the kicks dispersion. When it becomes sufficiently large, the disorder created by the flow in the kick bath is transmitted to the spin ensemble which loses its coherence and evolves to the quantum microcanonical distribution.

\subsection{Lyapunov exponent analysis}
In order to analyse the horizon of coherence with a more quantitative viewpoint, we consider more general hyperbolic automorphisms of the torus: $\Phi = \left(\begin{array}{cc} 1 & 1 \\ x & x \pm 1 \end{array} \right)$ for $x \in \mathbb R^*$. Since $|\det \Phi| = 1$ the flow is conservative, and since the eigenvalues of $\Phi$ are not root of 1, the flow is ergodic and chaotic \cite{katok,coudene,benoist}. Let $\lambda_\pm \in \mathbb R$ be the eigenvalues of $\Phi$.
\begin{equation}
\Phi e_\pm = \lambda_\pm e_\pm
\end{equation}
with $e_\pm \in \mathbb T^2$ ($\|e_\pm\|=1$) and $|\lambda_+|>1$ ($|\lambda_+\lambda_-|=1$). $e_+$ indicates the unstable direction of the flow on $\mathbb T^2$ whereas $e_-$ indicates the stable direction. $\ln |\lambda_\pm|$ are the Lyapunov exponents of the dynamical system.\\
In order to simplify the discussion, we consider an initial distribution of first kicks $\{(\lambda^{(i)}_0,\varphi^{(i)}_0)\}_{i=0,...,N}$ randomly chosen into the square with base point $(\lambda_*,\varphi_*)$ and sides $d_0 e_+$ and $d_0 e_-$. After $n$ iterations, the maximal separation of the kicks is (for $n$ sufficiently large)
\begin{eqnarray}
d_n & = &\left\|\Phi^n \left(\left(\begin{array}{c} \lambda_* \\ \varphi_* \end{array} \right) + d_0 e_+ + d_0e_- \right) - \Phi^n \left(\begin{array}{c} \lambda_* \\ \varphi_* \end{array} \right)\right\| \nonumber \\
& = & \left\|\Phi^n \left(d_0 e_+ + d_0e_- \right)\right\| \nonumber \\
& = & \| \lambda_+^n e_+ + \lambda_-^n e_-\| d_0 \nonumber \\
& = & \sqrt{\lambda_+^{2n} + \lambda_-^{2n}} d_0 \nonumber \\
& \simeq & |\lambda_+|^n d_0
\end{eqnarray}
Let $d_\Box$ be the length of a classical microstate of an equipartition of $\mathbb T^2$ ($\mathbb T^2$ is covered by a set of disjoint cells of dimensions $d_\Box \times d_\Box$ which constitute the classical microstates). Disorder appears in the kick bath when the SIC forbids predictions on the orbits with a sufficiently accuracy (the dispersion of the kicks becomes too large). For a classical dynamical system this minimal dispersion length is fixed to be the length of a classical microstate. The kick bath is then caracterized by a classical horizon of predictability equal to
\begin{equation}
n_\Box = \frac{\ln d_\Box-\ln d_0}{\ln|\lambda_+|}
\end{equation}
($\forall n > n_\Box$ we have $d_n > d_\Box$). For an initial distribution of first kicks randomly chosen in the square $[\lambda_*,\lambda_*+d_0] \times [\varphi_*,\varphi_*+d_0]$ the previous formula is not completely satisfactory. Indeed the projections of this distribution on the stable and unstable axis are not uniform distributions of support length equal to $d_0$. Let $\gamma = \arctan \frac{e_+^{(\varphi)}}{e_+^{(\lambda)}}$ be the angle between $e_+$ and the $\lambda$-axis of $\mathbb T^2$. The dispersion of the projection of the initial distribution on the unstable axis is approximately $d_0/\sin \gamma$. The horizon of predictability of the kick bath is then
\begin{equation}
n_\Box = \frac{\ln d_\Box - \ln (d_0/\sin \gamma)}{\ln |\lambda_+|}
\end{equation}

\subsection{Entropy evolutions}
\begin{figure}
\includegraphics[width=8.5cm]{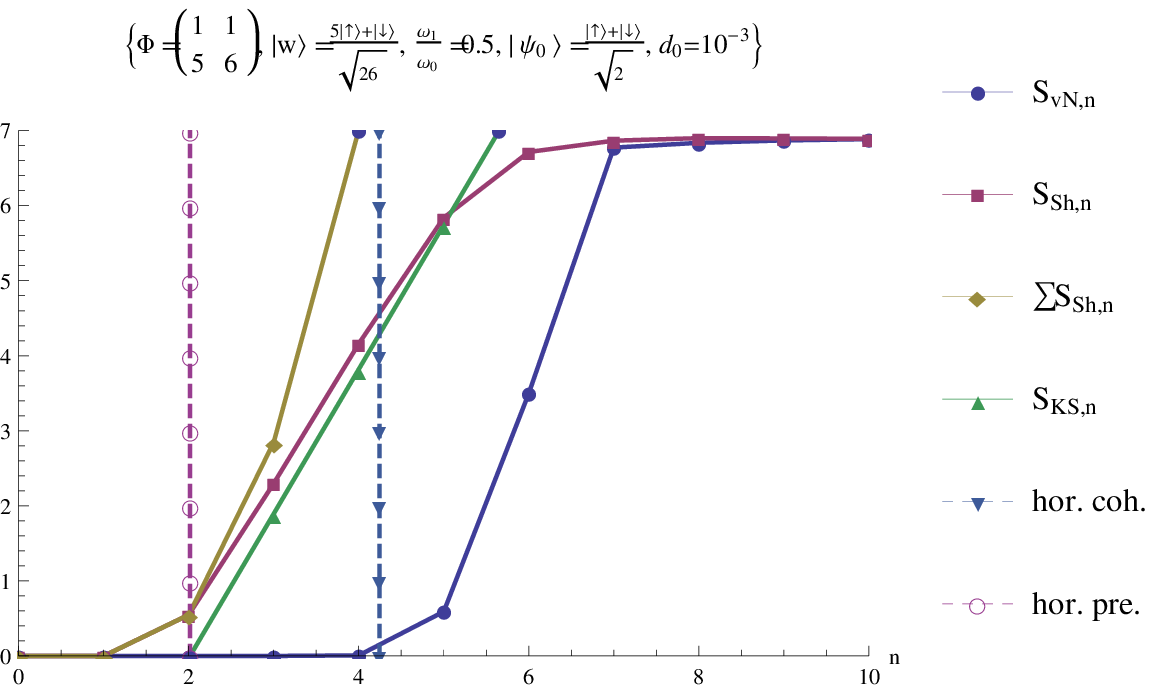}\\
\includegraphics[width=8.5cm]{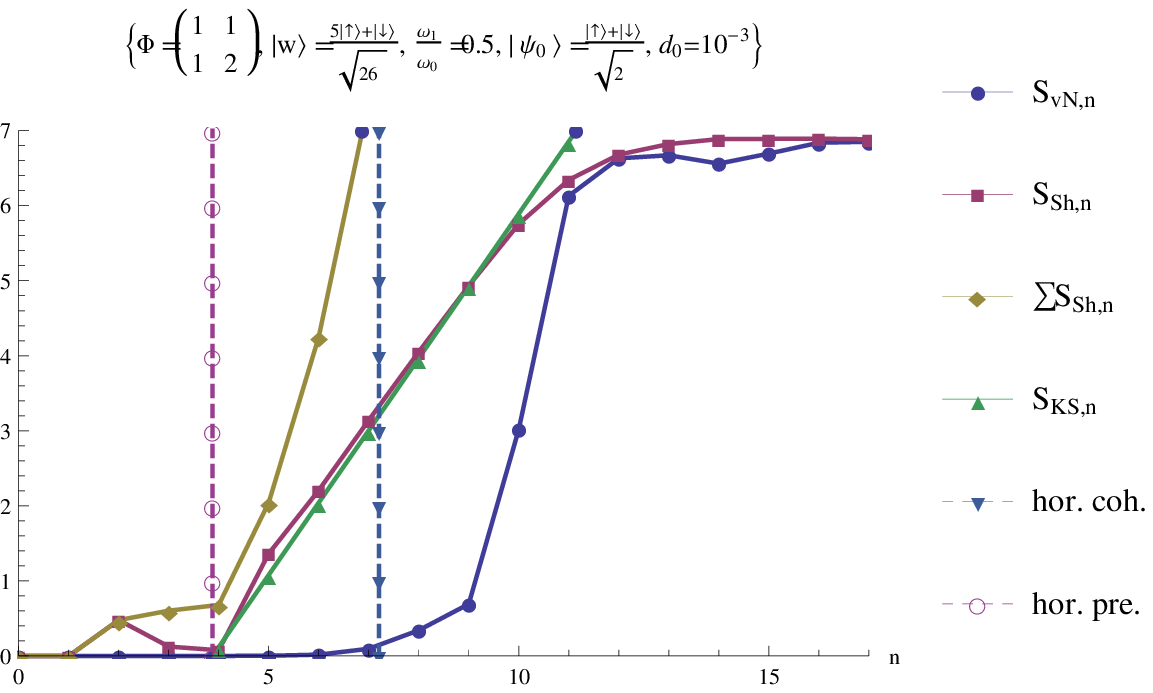}\\
\includegraphics[width=8.5cm]{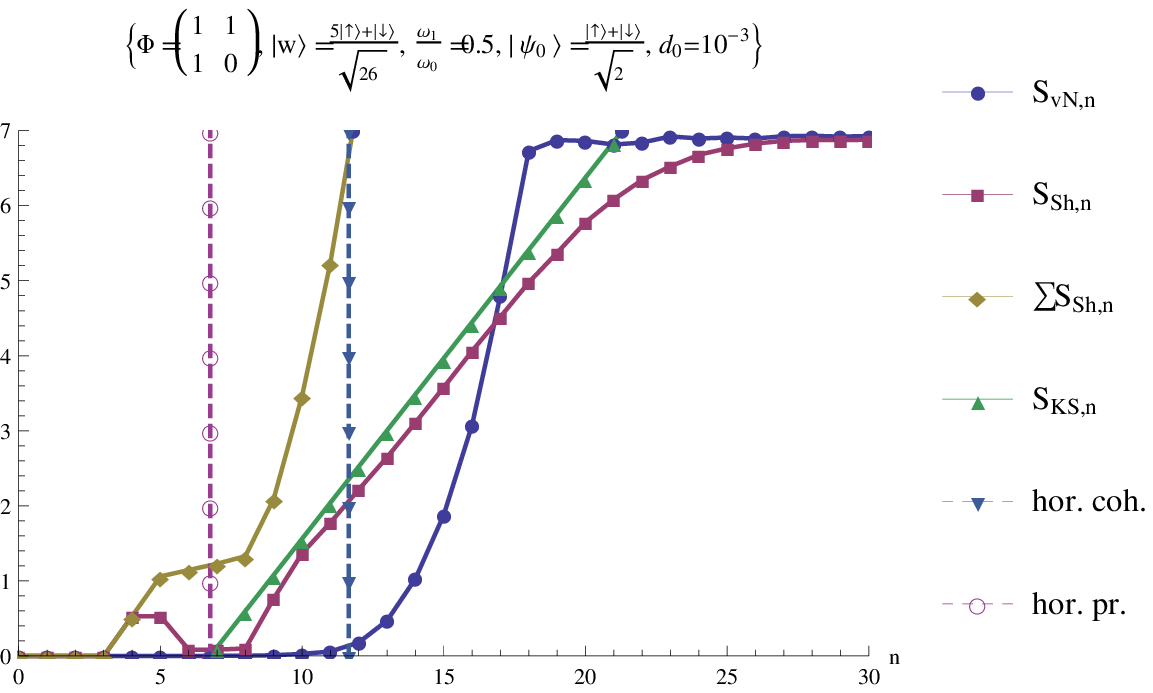}\\
\includegraphics[width=8.5cm]{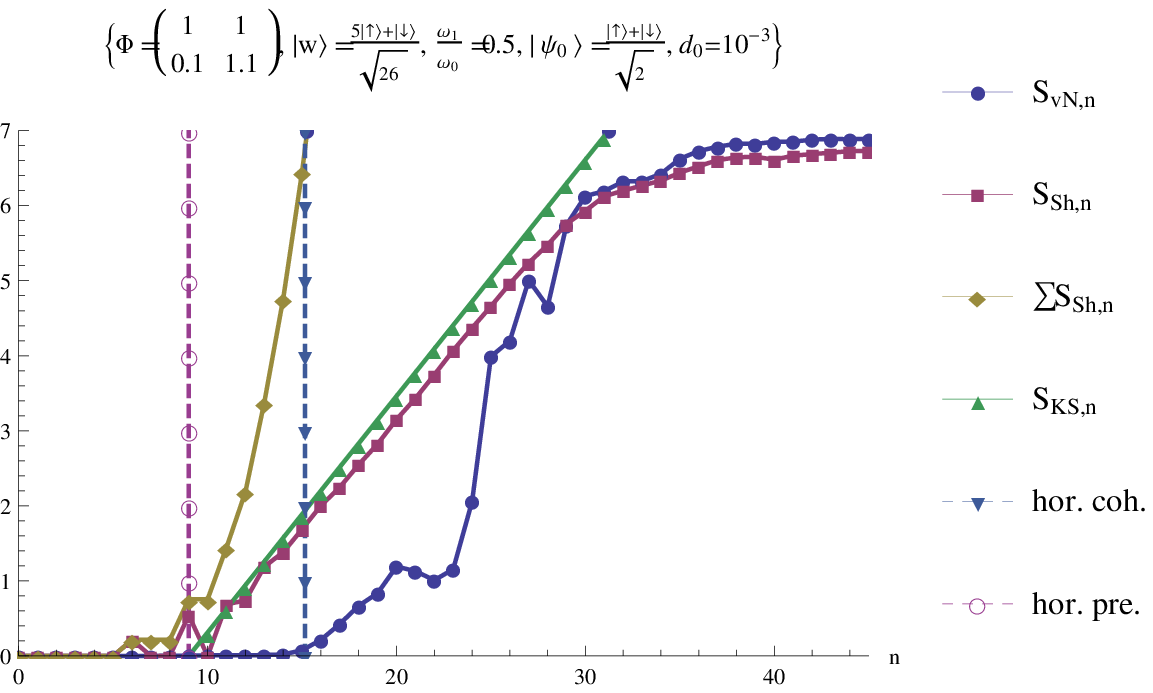}
\caption{\label{entropy} For different chaotic flows: von Neumann entropy of the spin ensemble, Shanon entropy of the kick bath, cumulated Shanon entropy of the kick bath and entropy of the kick bath predicted by the Kolmogorov-Sina\"i analysis. The horizon of predictability of the kick bath and the horizon of coherence of the spin ensemble are indicated by vertical dashed lines.}
\end{figure}

The main physical phenomenon in the system is related to the production and the transmission of disorder: the flow produces disorder in the kick bath due to its chaotic behavior (SIC), and this disorder is transmitted to the spin ensemble inducing decoherence. Now we want to analyse more precisely this phenomenon and we need to measure the disorder in the baths. Entropy functions are measures of disorder \cite{breuer,bengtsson}.\\
The disorder in the spin ensemble is measured by using the von Neumann entropy:
\begin{equation}
S_{vN,n} = - 10\ \tr \left(\rho_n \log \rho_n \right)
\end{equation}
where $\tr$ denotes the matricial trace, $\log$ denotes the matricial natural logarithm and the factor $10$ is arbitrary.\\
To define the disorder in the kick bath we need previously to choose classical microstates of the classical system. Let $X$ be the partition of $\mathbb T^2$ defined by the grid $\{i\frac{\pi}{64}\}_{i=0,...,128} \times \{j \frac{\pi}{64}\}_{j=0,...,128}$. A cell of $X$ constitutes one of the classical microstates for one kick train. We choose here the number of spins, and then the number of kick trains, equal to $N=1024$. The length of a microstate ($d_\Box =\frac{\pi}{64}$) is chosen to have a small probability that several kicks of an uniform distribution are simultaneously in the same microstate. The disorder of the kick bath is measured by using the Shanon entropy:
\begin{equation}
S_{Sh,n} = - \sum_{i,j} p_{ij,n} \ln p_{ij,n}
\end{equation}
where $p_{ij,n}$ is the fraction of kick trains which are in the microstate $(i,j)$ at the $n$-th iteration. The number of spins $N=1024$ and the arbitrary factor $10$ in the von Neumann entropy are chosen in order that the maximal entropies be equal: $\sup S_{vN} = \sup S_{Sh} = 10 \ln 2 = \ln 1024$. This permits a direct comparison of classical and quantum entropies without scale distortions.\\
Finally the production of disorder by the flow can be measured by the Kolmogorov-Sina\"i entropy (so called metric entropy or measure-theoretic entropy) \cite{katok,coudene,benoist}. Let $X_n = \bigvee_{p=0}^{n-1} \Phi^{-p}(X)$ be the partition of $\mathbb T^2$ refined by $\Phi$ with $\Phi^{-1}(X) = \{\Phi^{-1}(\sigma)\}_{\sigma \in X}$ and $X \vee Y = \{\sigma \cap \varsigma\}_{\sigma \in X; \varsigma \in Y}$. Let $\mu$ be the Haar probability measure on $\mathbb T^2$. The Kolmogorov-Sina\"i entropy of the flow is defined to be
\begin{equation}
h_\mu(\Phi) = - \sup_X \lim_{n \to + \infty} \frac{1}{n} \sum_{\sigma \in X_n} \mu(\sigma) \ln \mu(\sigma)
\end{equation}
It is the average disorder produced by $\Phi$ at each iteration. For hyperbolic automorphisms of $\mathbb T^2$ we have \cite{katok,coudene,benoist}:
\begin{equation}
h_\mu(\Phi) = \ln |\lambda_+|
\end{equation}
Entropy production starts in the kick bath at $n_\Box$ with a production rate equal to $\ln|\lambda_+|$, the entropy of the kick bath must be theoretically estimated to be
\begin{equation}
S_{KS,n} = \begin{cases} 0 & \text{if } n\leq n_\Box \\ (n-n_\Box) \ln|\lambda_+| & \text{if } n\geq n_\Box \end{cases}
\end{equation}
(we can suppose that $S_{Sh,n} \simeq S_{KS,n}$). The effect of the disorder in the kick bath is cumulative on the spin ensemble. Even if the entropy of the kick bath rests small, at each kick it induces a small increase of disorder of the spin ensemble. The disorder in the spin ensemble increases even if the disorder in the kick bath does not increase significantly. The entropy of the spin ensemble increases then if the cumulated entropy of the kick bath exceeds a threshold value. Numerical simulations show that this threshold value is the maximal entropy $S_{max} = \sup S_{Sh} = 10 \ln 2$. The horizon of coherence $n_\star$ is then such that
\begin{equation}
\label{def_nstar}
\sum_{n=0}^{n_\star} S_{Sh,n} = S_{max}
\end{equation}
Since $S_{Sh,n} \simeq S_{KS,n}$ we have
\begin{equation}
\sum_{n=n_\Box}^{n_\star} (n-n_\Box) \ln |\lambda_+| = S_{max}
\end{equation}
and then
\begin{equation}
\frac{(n_\star-n_\Box)(n_\star-n_\Box+1)}{2} \ln |\lambda_+| = S_{max}
\end{equation}
Finally, the theoretical horizon of coherence of the spin ensemble is
\begin{equation}
\label{nstar_th}
n_\star = n_\Box + \frac{1}{2} \sqrt{1+\frac{8 S_{max}}{\ln|\lambda_+|}} - \frac{1}{2}
\end{equation}
Fig. \ref{entropy} shows $S_{vN,n}$, $S_{Sh,n}$, $\sum_{p=0}^n S_{Sh,p}$ and $S_{KS,n}$ for different chaotic flows.
Fig. \ref{arnold_cat} shows that the formula (\ref{nstar_th}) is consistent with the numerical results concerning the coherence and the populations of the spin ensemble.\\
 Remark: in some cases, particularly when $x$ is small ($|x|<1$), the theoretical formula (\ref{nstar_th}) does not correspond exactly with numerical results. This is caused by small failures of the Kolmogorov-Sina\"i predictions for the classical entropy. Small variations of $S_{Sh,n}$ can occur before $n_\Box$ increasing the cumulated entropy. Moreover for $x$ small (weakly chaotic systems) the entropy production rate can be smaller than $\ln |\lambda_+|$ (since this value issued from the Kolmogorov-Sina\"i analysis, is an average value for very large $n$ and for optimal partitions $X$). Nevertheless the formula (\ref{def_nstar}) is always valid (it is the approximation $S_{Sh,n} \simeq S_{KS,n}$ which can present small failures).

\section{Summary and conclusion}
A spin ensemble controlled by disturbed kick train is subject to different dynamical effects: decoherence, relaxation, population oscillations, population jumps and horizons of coherence (for the chaotic case). These different phenomena can be controlled to a certain extend by adjust the system parameters as summarizes table \ref{control}.
\begin{table}
\caption{\label{control} Summary of the roles of the different parameters.}
\begin{tabular}{|c|l|}
\hline
\it paramaters  & \it effects on dynamical processes\\
\hline \hline
$ \frac{\omega_1}{\omega_0} \nearrow $ & decoherence efficiency $\nearrow$ \\
& relaxation efficiency  $\nearrow$ \\
& population oscillation frequency $\nearrow$\\
& population jump effect $\searrow$ \\
\hline
$ \vartheta \to 0$ or $\frac{\pi}{2}$ & decoherence efficiency $\nearrow$ \\
& relaxation efficiency $\searrow$ \\
& population oscillation amplitude $\searrow$ \\
& population jump effect $\searrow$ \\
\hline
$d_0 \nearrow$ & decoherence effect $\nearrow$ \\
& relaxation effect $\nearrow$ \\
& population oscillation effect $\searrow$ \\
& population jump effect $\nearrow$ \\
& horizon of coherence $\searrow$ (chaotic cases)\\
\hline
$\sigma \nearrow$ & population oscillation damping $\nearrow$ \\
(Markovian cases) & \\
\hline
$ \ln |\lambda_+| \nearrow$ & horizon of coherence $\searrow$ \\
(chaotic cases) & \\
\hline
\end{tabular}
\end{table}
The different dynamical effects could also permit to distinguish the different classical baths by studying the coherence and the populations of the quantum bath as summarizes table \ref{baths_behav}.
\begin{table}
\caption{\label{baths_behav} Summary of the distingushing behaviors of the spin ensemble with respect to the kick baths.}
\begin{tabular}{|l|l|}
\hline
\it classical baths & \it specific behavior of the quantum bath \\
\hline \hline
stationary bath & no or incomplete decoherence \\
& no or incomplete relaxation \\
\hline
drifting bath & no or strongly fluctuating decoherence \\
& no or strongly fluctuating relaxation \\
\hline
microcanonical bath & rapid complete decoherence \\
& rapid complete relaxation \\
\hline
Markovian bath & neat damped population oscillations \\
\hline
chaotic bath & horizon of coherence \\
\hline
\end{tabular}
\end{table}
Chaotic kick baths are particularly interesting since they present two distinct behaviors. At short-term the spin ensemble presents a behavior without decoherence and relaxation processes whereas at long-term decoherence and relaxation processes dominate the dynamics. Spin ensembles chaotically kicked present then a \textit{horizon of coherence} which is a quantum analogue to the horizons of predictability of chaotic classical systems and which is a direct consequence of the sensitivity to initial conditions into the kick bath. An interesting question could be to know if similar behaviors occur for quantum baths in contact with other kinds of classical baths.

\end{document}